\documentclass[preprint,showpacs,preprintnumbers,amsmath,amssymb]{revtex4}
\usepackage{epsfig}
\usepackage{graphicx}
\usepackage{dcolumn}
\usepackage{bm}
\def\lsim{\mathrel {\vcenter {\baselineskip 0pt \kern 0pt
    \hbox{$&lt;$} \kern 0pt \hbox{$\sim$} }}}
\def\gsim{\mathrel {\vcenter {\baselineskip 0pt \kern 0pt
    \hbox{$&gt;$} \kern 0pt \hbox{$\sim$} }}}

\newcommand{\U}{{\cal {U}}}

\begin{document}

\title{Unparticle Effects on Unitarity Constraints \\from Higgs Boson Scattering}

\author{Xiao-Gang He and Chung-Cheng Wen }
\affiliation{Department of Physics and Center for
Theoretical Sciences, National Taiwan University, Taipei, Taiwan, R.O.C.}

\date{\today}

\begin{abstract}
We study the effects of two-body Higgs boson scattering by
exchanging unparticles. The unparticle contribution can change the
standard model prediction for two-body Higgs boson scattering
partial wave amplitude significantly leading to modification of
the unitarity constraint on the standard model Higgs boson mass. For
unparticle dimension $d_\U$ between 1 and 2, the unitarity constraint on
Higgs boson mass can be larger than that in the SM. Information on
unparticle interaction can also be obtained.
\end{abstract}


\maketitle

Since the seminal work of Georgi on unparticle
physics\cite{Georgi:2007ek} last year, the study of unparticle
effects has drown a lot of attentions\cite{Georgi:2007ek,0704,0705,0706,0707,0708,0709,0710,0711,0801,Fox:2007sy,Chen:2007qr}.
The concept of unparticle
\cite{Georgi:2007ek} stems from the observation that certain high
energy theory with a nontrivial infrared fixed-point at some scale
$\Lambda_{\U}$ may develop a scale-invariant degree of freedom
below the scale. The kinematics is determined by its scaling
dimension $d_\U$ under scale transformations. The unparticle must
interact with Standard Model (SM) particles to be physically relevant.
Even though at
present the detailed dynamics of how unparticle interacts with SM
 particles is not known, these interactions can
well be described in effective field theory. In this approach the
interactions are parameterized in the following way\cite{Georgi:2007ek}
\begin{eqnarray} \lambda \Lambda_{\cal{U}} ^{4-d_{SM} - d_\U}
O_{SM} O_{\cal{U}},
\end{eqnarray}
where $O_{SM}$ is composed of the SM fields, and $O_\U$
is an unparticle operator.

There has been a burst of activities on various aspects of
unparticle physics from phenomenology to theoretical
issues\cite{Georgi:2007ek,0704,0705,0706,0707,0708,0709,0710,0711,0801,Fox:2007sy,Chen:2007qr}.
Some of the major tasks of phenomenological study are to search
for new signals and effects in various physical processes, and to
determine (constrain) the unparticle scale and also unparticle
dimension $d_\U$. In this work we study unparticle interaction
effects on unitarity constrains from two-body Higgs boson
scattering using partial wave analysis. We find
that the unparticle contribution to the scattering partial wave
amplitude can be significant which affect the unitarity constraint
on the Higgs boson mass. For $d_\U$ between 1 and 2, the
unparticle contribution can relax the upper bound for Higgs boson
mass.

Partial wave analysis of scattering processes is one of the often
used methods to constrain unknown parameters in a theory. The
unitarity constraint on Higgs boson mass from two-body Higgs boson
scattering in the SM\cite{Lee:1977eg}, and constraint on
extra dimension scale\cite{He:1999qv} are some of the interesting
examples.  A scattering amplitude $\mathcal{M}$ for a given
process can be decomposed into partial wave amplitude according to
angular momentum $\vec J$ as
\begin{eqnarray}
\mathcal{M} = \frac{1}{k}\sum{a_{J}(2J+1)P_{J}(\cos\theta)}.
\end{eqnarray}
The unitarity condition is referred to the condition that the
magnitude for each of the partial wave amplitude $|a_J|$ should
not be too large. There are many discussions of how to implement
unitarity condition to constrain new physics\cite{dawson}.
We will use a weak condition
$|a_J|<1$ for $J=0$ and work with the tree level amplitude to show
how interesting constraints on unparticle interactions and Higgs
boson mass can be obtained.

Potentially large contribution to the two-body Higgs boson
scattering may come from the following lowest dimension operator involving a
scalar unparticle and SM Higgs field\cite{Fox:2007sy,Chen:2007qr}
\begin{eqnarray}
O_{hh} = \lambda_{hh} \Lambda_{\cal{U}}^{2-d_\U}H^\dagger H
O_{\cal{U}},
\end{eqnarray}
where $H = (h^+, (v + h + i I)/\sqrt{2})$ is the SM Higgs doublet.
The $h^+$ and $I$ are the fields ``eaten'' by $W$ and $Z$, and $h$
is a physical Higgs. The parameter $\lambda_{hh}$ is real.

There are $s$, $t$ and $u$ channel contributions from the above
effective operator to the two-body Higgs boson scattering amplitude as
shown in Fig. 1. We obtain the scattering amplitude as
\begin{eqnarray}
\mathcal{M}^{un}(h h \to h h) &=&(\lambda_{hh}
\Lambda^{2-d_{\cal{U}}}_{\cal{U}})^2
\frac{A_{d_{\cal{U}}}}{2\sin(\pi d_{\cal{U}})}
[\frac{1}{(-s)^{d_{\cal{U}}}}+\frac{1}{(-t)^{d_{\cal{U}}}}+\frac{1}{(-u)^{d_{\cal{U}}}}].
\end{eqnarray}
In obtaining the above expression, we have used the scalar
unparticle propagator $(iA_{d_\U}/2\sin(\pi d_\U))
(1/(-p^2)^{2-d_\U})$. The factor $A_{d_{\cal{U}}}$ is normalized
as $A_{d_\U} =
(16\pi^{5/2}/(2\pi)^{2d_\U})\Gamma(d_\U+1/2)/(\Gamma(d_\U
-1)\Gamma(2d_\U))$ following ref.\cite{Georgi:2007ek}.

\begin{figure}
  \graphicspath{{image3/}}
  \includegraphics[width=10cm]{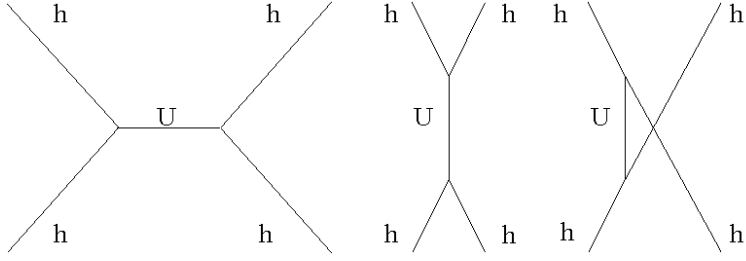}
  \caption{Feynman diagrams for two-body Higgs boson scattering by exchanging an unparticle
  in $s$, $t$, and $u$ channels. }\label{Fig.1}
\end{figure}

Using the above scattering amplitude, the J=0 component in the
partial wave expansion $a^{un}_0$ can be easily obtained
\begin{eqnarray}
a^{un}_{0}&=&\frac{1}{16\pi}(\frac{4\vec{p}^2}{s})^{1/2}
\frac{1}{s-4m^{2}_{h}} \int_{-(s-4m^{2}_{h})}^{0} \mathcal{M}^{un}
dt\nonumber\\
& =&\frac{1}{16\pi}\lambda_{hh}^2 \left ({\sqrt{s}\over
\Lambda_\U}\right
)^{2d_{\cal{U}}-4}\frac{A_{d_{\cal{U}}}}{2\sin(\pi
d_{\cal{U}})}\sqrt{1-\frac{4m^{2}_h}{s}}[e^{-i\pi
d_{\cal{U}}}+\frac{2}{d_{\cal{U}}-1}(1-{4m^{2}_{h}\over
s})^{d_{\cal{U}}-2}],\label{am}
\end{eqnarray}
where $\vec p = \sqrt{1-4m^2_h/s}$ is the Higgs boson momentum in
the center of mass frame.

As for any other processes involving unparticle propagator, there
is a $\sin(\pi d_{\cal{U}})$ factor in the denominator which has
poles at integer $d_\U$ and makes $a^{un}_{0}$ to diverge. In the
first term in eq.(\ref{am}), the pole at $d_\U = 1$ is cancelled
by a zero in $A_{\cal{U}}$. However, the second term will blow
off. Therefore integer numbers are forbidden. Also the factor
$A_{d_{\cal{U}}}$ decreases quickly as $d_\U$ increases, therefore for large
$d_\U$ the unparticle contribution is suppress.

For a complete analysis, one also needs to include the SM
contribution where the $J=0$ partial wave amplitude is given
by\cite{Lee:1977eg}
\begin{eqnarray}
a^{SM}_0=\frac{G_Fm^2_h}{8\sqrt{2}\pi}\sqrt{1-\frac{4m^2_h}{s}}
[3+\frac{9m^2_h}{s-m^2_h}-\frac{18m^2_h}{s-4m^2_h}\ln(\frac{s}{m^2_h}-3)].
\end{eqnarray}
With this contribution included, the weak unitarity condition
becomes
\begin{eqnarray}
|a^T_0| = |a^{SM}_0 + a^{un}_0| < 1.
\end{eqnarray}

There is an imaginary part from unparticle contribution to $a_0$
due to the s-channel unparticle exchange in Fig. 1 with
$Im a_0^{un} = -(1/32\pi)r^{2d_\U - 4}A_{d_\U}
\sqrt{1-4m^2_h/s}$. Here $r = (\lambda_{hh})^{1/(d_\U - 2)}
\sqrt{s}/\Lambda_\U$. Since the SM contribution is real, the
unitarity condition requires $|Im a_0| < 1$. One can, in
principle, obtain a constraint on the parameter $r$ as a function
of the unparticle dimension. We have analyzed this and found that
the constraints are weak. The combined effects of real and
imaginary parts can provide more interesting information which we study in the following.

In the limiting case of $s>> 4 m_h^2$, weak unitarity condition is
simply given by,
\begin{eqnarray}
|r
^{2d_{\cal{U}}-4}\frac{A_{d_{\cal{U}}}}{32\pi \sin(\pi
d_{\cal{U}})}[e^{-i\pi
d_{\cal{U}}}+\frac{2}{d_{\cal{U}}-1}]  + \frac{3G_Fm^2_h}{8\sqrt{2}\pi}|<1.
\end{eqnarray}

\begin{figure}
  \graphicspath{{image3/}}
  \includegraphics[width=6.5cm]{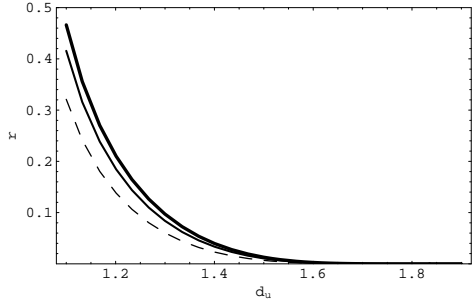}\\
  \caption{Lower bound on $r$ as a function of $d_\U$ with different Higgs masses,
  115 (solid line), 500 (lighter solid line), 1000 (dashed line) GeV in the limit $s >> 4 m^2_h$.}\label{Fig.2}
\end{figure}

Since the strength of the unparticle contribution to the partial
wave amplitude is a function of $r$, the unitarity condition may
provide information about $r$. We plot, in Fig. 2, $r$ as a
function of $d_\U$ for several representative Higgs boson masses for
$d_\U$ in the range between 1 and 2. Note that the unitarity bound
gives a lower bound for $r$ because $2d_\U -4 <0$. This reflects
the fact that the interaction of Higgs boson with unparticle defined by
operator $O_{hh}$ does not decouple in the limit $\Lambda_\U$ goes
to infinite. Since $r \sim \sqrt{s}/\Lambda_\U$, naively, for
$d_\U$ smaller than 2, small $s$ is ruled out. However, one must
keep in mind that $s
> 4 m^2_h$ must be satisfied, $s$ smaller than $4 m^2_h$ is not constrained by unitarity
condition.   For $d > 2$, the unitarity bound gives a upper bound
for $r$. In this case, in the large $\Lambda_\U$ limit, the
interaction of Higgs boson and unparticle decouples.

If the unparticle scale $\Lambda_\U$ is known from some
theoretical considerations, one can use the weak unitarity condition
to constrain the energy scale $\sqrt{s}$ with which one can
reliably (satisfying the weak unitarity condition) use the
operator $O_{hh}$ for calculations. We have carried out a study keeping
$4m^2_h/s$ term in the expression for $a^T_0$. For
$d_\U$ larger than 2, the unitarity condition enables one to obtain an upper bound 
for $s$ since the leading $s$ dependence is $s^{d_\U-2}$. $s$ cannot be too large in order not to violate 
the unitarity condition, 
but numerically it is way above 10 TeV or any near future collider energies, such as
LHC and ILC. For $d_\U$ between 1 and 2, the unitarity condition puts a low bound for
$s$. Since the leading scale $\Lambda_\U$ and $s$ dependence of $a^T_0$ is $(s/\Lambda_\U^2)^{d_\U-2}$, 
the unitarity condition gives a lower bound for $s$. A smaller $\Lambda_\U$ corresponds to a larger $s$.
Numerically we find that for lower values of $\Lambda_\U$ (less than 1 TeV), $s$ larger than the threshold is
all allowed. But for larger $\Lambda_\U$, for example 10 TeV, there are regions with $d_\U $ close to 1 violate 
unitarity condition for $s$ above the threshold. Note also that near the threshold, the second term in eq.(5) become very large and therefore
$|a^T_0|$, if $d_\U$ is smaller than 1.5. One should not use value for s too close to the
threshold if $d_\U$ is less than 1.5.

We find that the weak
unitarity condition is satisfied for $s$ significantly larger than the threshold
of producing two Higgs bosons for $d_\U$ between 1 and 2. 

We now discuss unparticle effects on unitarity constraint on Higgs
boson mass $m_h$. Without unparticle contribution, in the limit $s
>> 4 m^2_h$, the weak unitary condition implies that the Higgs
boson mass must be smaller than $8\sqrt{2}\pi/3 G_F = 1010$ GeV.
With unparticle contributions, the constraint on Higgs mass can be
modified dramatically since the real part of the unparticle
contribution can have either signs relative to the SM contribution
depending on the unparticle dimension $d_\U$. For example, for
$d_\U$ between 1 and 2, the real part of $a^{un}$ is negative
making the constraint on Higgs mass looser compared the one for
SM. For $d_\U$ between 2 and 3, $Re(a^{un})$ is positive, the
constraint on Higgs mass becomes tighter. Since for large $d_\U$,
there is a suppression from $A_{d_\U}$, the constraint on relevant
parameters are weak. We will concentrate on $d_\U$ between 1 and
2.

\begin{figure}
  \graphicspath{{image3/}}
  \includegraphics[width=4.8cm]{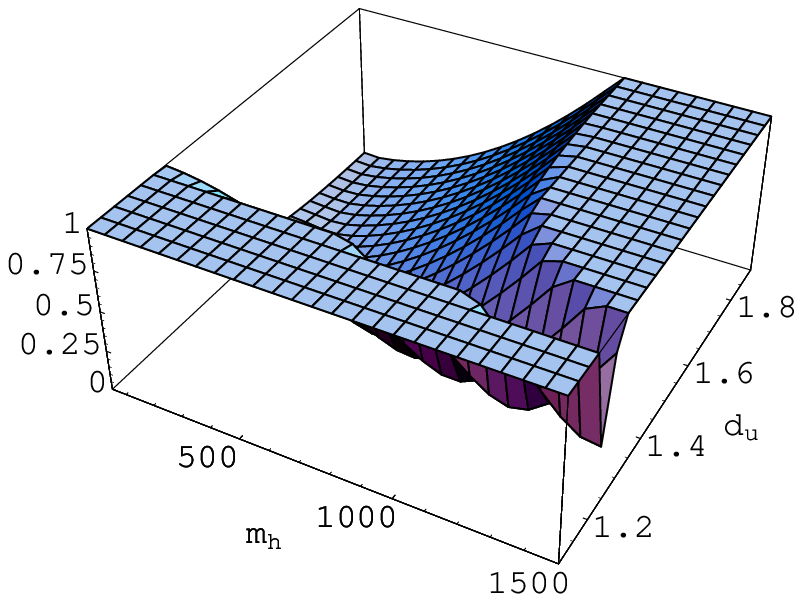}
  \includegraphics[width=4.8cm]{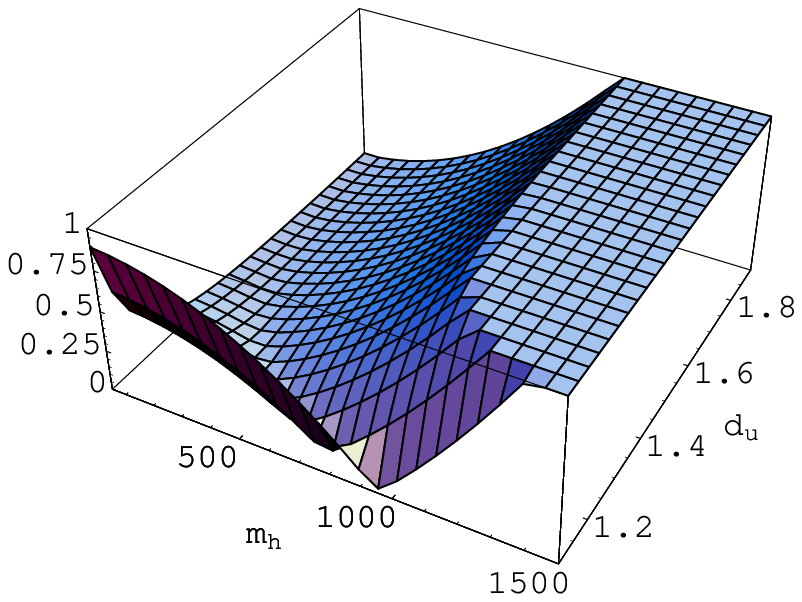}
  \includegraphics[width=4.8cm]{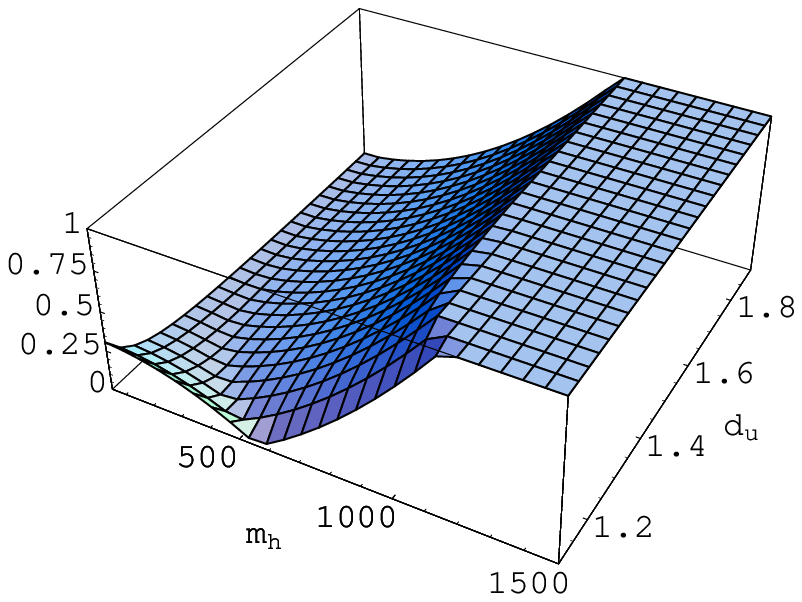}
 \caption{$|a^0_t|$ (vertical axis)
 as a function of Higgs mass( in GeV) and $d_\U$ with $r=0.1, 0.5$ and $1$.}\label{Fig.3.1}
\end{figure}

\begin{figure}
  \graphicspath{{image3/}}
  \includegraphics[width=7cm]{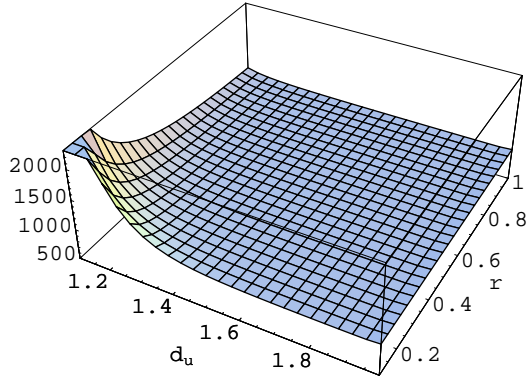}
 \caption{The upper bound of Higgs mass $m_h$ in GeV (vertical axis) as a function of $d_\U$ and $r$.}\label{Fig.4.1}
\end{figure}

In Fig. 3, we show $|a^T_0|$ as functions of the Higgs boson mass
$m_h$ and the unparticle dimension $d_\U$ for several finite
values of $r$ in the limit $s>> 4 m^2_h$. With a low value for
$r$, the allowed region in $m_h$ and $d_\U$ space is more restrictive
than those for larger $r$. This is because that for smaller $r$, $a_0^{un}$
becomes larger as $d_\U$ decreases. To satisfy the unitarity constraint,
a large cancelation from the SM contribution is needed and resulting in a larger
Higgs boson mass.
Fixing $|a^T_0| = 1$, one
can solve an upper bound for the Higgs mass $m_h$ as a function of
$d_\U$ and r. In Fig. 4 we show this upper bound. We see more clearly that for smaller $r$ and $d_\U$
much larger Higgs boson mass compared with SM unitarity bound is allowed.
When $d_\U$ and $r$ become larger, the
unparticle effects decreases. The unitarity bound on Higgs boson
mass quickly, from above, reaches the SM one.

In summary we have studied unparticle effects on the unitarity
constraints from two-body Higgs boson scattering process. We find
that the unparticle contribution to the scattering partial wave
amplitude can be significant which affect the unitarity constraint
on the Higgs boson mass. For $d_\U$ between 1 and 2, the
unparticle contribution can relax the upper bound for Higgs boson
mass. For $d_\U$
smaller than 1.3 and $r$ smaller than 0.4, the allowed Higgs boson mass can be
much larger than that in the SM.


\begin{references}
\bibitem{Georgi:2007ek}
  H.~Georgi,
  Phys.\ Rev.\ Lett.\  {\bf 98}, 221601 (2007);
 H.~Georgi,
  Phys.\ Lett.\  B {\bf 650}, 275 (2007)
  [arXiv:0704.2457 [hep-ph]].

  \bibitem{0704}
  K.~Cheung, W.~Y.~Keung and T.~C.~Yuan,
  Phys.\ Rev.\ Lett.\  {\bf 99}, 051803 (2007)
  [arXiv:0704.2588 [hep-ph]];
  M.~Luo and G.~Zhu,
  arXiv:0704.3532 [hep-ph].

\bibitem{0705}
  C.~H.~Chen and C.~Q.~Geng,
  arXiv:0705.0689 [hep-ph];
  G.~J.~Ding and M.~L.~Yan,
  Phys.\ Rev.\  D {\bf 76}, 075005 (2007)
  [arXiv:0705.0794 [hep-ph]];
  Y.~Liao,
  Phys.\ Rev.\  D {\bf 76}, 056006 (2007)
  [arXiv:0705.0837 [hep-ph]];
  T.~M.~Aliev, A.~S.~Cornell and N.~Gaur,
  Phys.\ Lett.\  B {\bf 657}, 77 (2007)
  [arXiv:0705.1326 [hep-ph]];
  X.~Q.~Li and Z.~T.~Wei,
  Phys.\ Lett.\  B {\bf 651}, 380 (2007)
  [arXiv:0705.1821 [hep-ph]];
  C.~D.~Lu, W.~Wang and Y.~M.~Wang,
  Phys.\ Rev.\  D {\bf 76}, 077701 (2007)
  [arXiv:0705.2909 [hep-ph]];
  M.~A.~Stephanov,
  Phys.\ Rev.\  D {\bf 76}, 035008 (2007)
  [arXiv:0705.3049 [hep-ph]];
  N.~Greiner,
  Phys.\ Lett.\  B {\bf 653}, 75 (2007)
  [arXiv:0705.3518 [hep-ph]];
  H.~Davoudiasl,
  Phys.\ Rev.\ Lett.\  {\bf 99}, 141301 (2007)
  [arXiv:0705.3636 [hep-ph]];
  D.~Choudhury, D.~K.~Ghosh and Mamta,
  arXiv:0705.3637 [hep-ph];
  T.~M.~Aliev, A.~S.~Cornell and N.~Gaur,
  JHEP {\bf 0707}, 072 (2007)
  [arXiv:0705.4542 [hep-ph]];
  P.~Mathews and V.~Ravindran,
  Phys.\ Lett.\  B {\bf 657}, 198 (2007)
  [arXiv:0705.4599 [hep-ph]].



 \bibitem{0706}
  S.~Zhou,
  arXiv:0706.0302 [hep-ph];
  G.~J.~Ding and M.~L.~Yan,
  arXiv:0706.0325 [hep-ph];
  C.~H.~Chen and C.~Q.~Geng,
  Phys.\ Rev.\  D {\bf 76}, 036007 (2007)
  [arXiv:0706.0850 [hep-ph]];
  Y.~Liao and J.~Y.~Liu,
  arXiv:0706.1284 [hep-ph];
  P.~Ball,
  M.~Bander, J.~L.~Feng, A.~Rajaraman and Y.~Shirman,
  arXiv:0706.2677 [hep-ph];
  T.~G.~Rizzo,
  JHEP {\bf 0710}, 044 (2007)
  [arXiv:0706.3025 [hep-ph]];
  K.~Cheung, W.~Y.~Keung and T.~C.~Yuan,
  Phys.\ Rev.\  D {\bf 76}, 055003 (2007)
  [arXiv:0706.3155 [hep-ph]].

  \bibitem{0707}
  S.~L.~Chen, X.~G.~He and H.~C.~Tsai,
   JHEP {\bf 0711}, 010 (2007)
  [arXiv:0707.0187 [hep-ph]];
  R.~Zwicky,
  arXiv:0707.0677 [hep-ph];
  T.~Kikuchi and N.~Okada,
  arXiv:0707.0893 [hep-ph];
  R.~Mohanta and A.~K.~Giri,
  Phys.\ Rev.\  D {\bf 76}, 075015 (2007)
  [arXiv:0707.1234 [hep-ph]];
  C.~S.~Huang and X.~H.~Wu,
  arXiv:0707.1268 [hep-ph];
  A.~Lenz,
  Phys.\ Rev.\  D {\bf 76}, 065006 (2007)
  [arXiv:0707.1535 [hep-ph]];
  D.~Choudhury and D.~K.~Ghosh,
  arXiv:0707.2074 [hep-ph];
  H.~Zhang, C.~S.~Li and Z.~Li,
  arXiv:0707.2132 [hep-ph];
  X.~Q.~Li, Y.~Liu and Z.~T.~Wei,
  arXiv:0707.2285 [hep-ph];
  Y.~Nakayama,
  Phys.\ Rev.\  D {\bf 76}, 105009 (2007)
  [arXiv:0707.2451 [hep-ph]];
  N.~G.~Deshpande, X.~G.~He and J.~Jiang,
  Phys.\ Lett.\  B {\bf 656}, 91 (2007)
  [arXiv:0707.2959 [hep-ph]];
  T.~A.~Ryttov and F.~Sannino,
  Phys.\ Rev.\  D {\bf 76}, 105004 (2007)
  [arXiv:0707.3166 [hep-th]];
  R.~Mohanta and A.~K.~Giri,
  Phys.\ Rev.\  D {\bf 76}, 057701 (2007)
  [arXiv:0707.3308 [hep-ph]];
  A.~Delgado, J.~R.~Espinosa and M.~Quiros,
  JHEP {\bf 0710}, 094 (2007)
  [arXiv:0707.4309 [hep-ph]].

  \bibitem{0708}
  M.~Neubert,
  arXiv:0708.0036 [hep-ph];
  M.~x.~Luo, W.~Wu and G.~h.~Zhu,
  arXiv:0708.0671 [hep-ph];
  S.~Hannestad, G.~Raffelt and Y.~Y.~Y.~Wong,
  arXiv:0708.1404 [hep-ph];
  N.~G.~Deshpande, S.~D.~H.~Hsu and J.~Jiang,
  arXiv:0708.2735 [hep-ph];
  P.~K.~Das,
  arXiv:0708.2812 [hep-ph];
  G.~Bhattacharyya, D.~Choudhury and D.~K.~Ghosh,
  Phys.\ Lett.\  B {\bf 655}, 261 (2007)
  [arXiv:0708.2835 [hep-ph]];
  Y.~Liao,
  arXiv:0708.3327 [hep-ph];
  D.~Majumdar,
  arXiv:0708.3485 [hep-ph];
  A.~T.~Alan and N.~K.~Pak,
  arXiv:0708.3802 [hep-ph];
  A.~Freitas and D.~Wyler,
  arXiv:0708.4339 [hep-ph];
  I.~Gogoladze, N.~Okada and Q.~Shafi,
  arXiv:0708.4405 [hep-ph].

  \bibitem{0709}
  C.~H.~Chen and C.~Q.~Geng,
  arXiv:0709.0235 [hep-ph];
  T.~i.~Hur, P.~Ko and X.~H.~Wu,
  arXiv:0709.0629 [hep-ph];
  L.~Anchordoqui and H.~Goldberg,
  arXiv:0709.0678 [hep-ph].
  S.~Majhi,
  arXiv:0709.1960 [hep-ph];
  J.~McDonald,
  arXiv:0709.2350 [hep-ph];
  M.~C.~Kumar, P.~Mathews, V.~Ravindran and A.~Tripathi,
  arXiv:0709.2478 [hep-ph];
  S.~Das, S.~Mohanty and K.~Rao,
  arXiv:0709.2583 [hep-ph];
  G.~j.~Ding and M.~L.~Yan,
  arXiv:0709.3435 [hep-ph];

  \bibitem{0710}
  A.~B.~Balantekin and K.~O.~Ozansoy,
  arXiv:0710.0028 [hep-ph];
  T.~M.~Aliev and M.~Savci,
  arXiv:0710.1505 [hep-ph];
  K.~Cheung, W.~Y.~Keung and T.~C.~Yuan,
  arXiv:0710.2230 [hep-ph];
  E.~O.~Iltan,
  arXiv:0710.2677 [hep-ph];
  S.~L.~Chen, X.~G.~He, X.~Q.~Li, H.~C.~Tsai and Z.~T.~Wei,
  arXiv:0710.3663 [hep-ph];
  I.~Lewis,
  arXiv:0710.4147 [hep-ph];
  A.~T.~Alan, N.~K.~Pak and A.~Senol,
  arXiv:0710.4239 [hep-ph];
  G.~L.~Alberghi, A.~Y.~Kamenshchik, A.~Tronconi, G.~P.~Vacca and G.~Venturi,
  arXiv:0710.4275 [hep-th];
  S.~L.~Chen, X.~G.~He, X.~P.~Hu and Y.~Liao,
  arXiv:0710.5129 [hep-ph];
  O.~Cakir and K.~O.~Ozansoy,
  arXiv:0710.5773 [hep-ph].

  \bibitem{0711}
  T.~Kikuchi and N.~Okada,
  arXiv:0711.1506 [hep-ph];
  I.~Sahin and B.~Sahin,
  arXiv:0711.1665 [hep-ph];
  E.~O.~Iltan,
  arXiv:0711.2744 [hep-ph];
  A.~T.~Alan,
  arXiv:0711.3272 [hep-ph];
  K.~Cheung, C.~S.~Li and T.~C.~Yuan,
  arXiv:0711.3361 [hep-ph];
  R.~Mohanta and A.~K.~Giri,
  arXiv:0711.3516 [hep-ph];
  J.~R.~Mureika,
  arXiv:0712.1786 [hep-ph].
  Y.~Wu and D.~X.~Zhang,
  arXiv:0712.3923 [hep-ph].

\bibitem{0801}
  T.~Kikuchi, N.~Okada and M.~Takeuchi,
  arXiv:0801.0018 [hep-ph];
  X.~G.~He and S.~Pakvasa,
  Phys.\ Lett.\  B {\bf 662}, 259 (2008)
  [arXiv:0801.0189 [hep-ph]];
  E.~O.~Iltan,
  arXiv:0801.0301 [hep-ph];
  C.~H.~Chen, C.~S.~Kim and Y.~W.~Yoon,
  arXiv:0801.0895 [hep-ph];
  V.~Bashiry,
  arXiv:0801.1490 [hep-ph];
  J.~L.~Feng, A.~Rajaraman and H.~Tu,
  arXiv:0801.1534 [hep-ph];
  K.~Cheung, T.~W.~Kephart, W.~Y.~Keung and T.~C.~Yuan,
  arXiv:0801.1762 [hep-ph];
  I.~Sahin,
  arXiv:0801.1838 ;
  V.~Barger, Y.~Gao, W.~Y.~Keung, D.~Marfatia and V.~N.~Senoguz,
  Phys.\ Lett.\  B {\bf 661}, 276 (2008)
  [arXiv:0801.3771 [hep-ph]];
  G.~J.~Ding and M.~L.~Yan,
  Phys.\ Rev.\  D {\bf 77}, 014005 (2008);
  M.~J.~Aslam and C.~D.~Lu,
  arXiv:0802.0739 [hep-ph];
  E.~O.~Iltan,
  arXiv:0802.1277 [hep-ph];
  B.~Sahin,
  arXiv:0802.1937 [hep-ph];
  I.~Sahin,
  arXiv:0802.2818 [hep-ph];
  A.~Hektor, Y.~Kajiyama and K.~Kannike,
  arXiv:0802.4015 [hep-ph];
  Y.~Gong and X.~Chen,
  arXiv:0803.3223 [astro-ph];
  M.~C.~Kumar, P.~Mathews, V.~Ravindran and A.~Tripathi,
  Phys.\ Rev.\  D {\bf 77}, 055013 (2008);
  F.~Sannino,
  arXiv:0804.0182 [hep-ph];
  E.~Iltan,
  arXiv:0804.2456 [hep-ph];
  M.~C.~Kumar, P.~Mathews, V.~Ravindran and A.~Tripathi,
  arXiv:0804.4054 [hep-ph];
  Y.~Liao,
  arXiv:0804.4033 [hep-ph].


\bibitem{Fox:2007sy}
  P.~J.~Fox, A.~Rajaraman and Y.~Shirman,
  Phys.\ Rev.\  D {\bf 76}, 075004 (2007)
  [arXiv:0705.3092 [hep-ph]].

\bibitem{Chen:2007qr}
  S.~L.~Chen and X.~G.~He,
  Phys.\ Rev.\  D {\bf 76}, 091702 (2007)
  [arXiv:0705.3946 [hep-ph]].

\bibitem{Lee:1977eg}
  B.~W.~Lee, C.~Quigg and H.~B.~Thacker,
  Phys.\ Rev.\  D {\bf 16}, 1519 (1977).


\bibitem{He:1999qv}
  X.~G.~He,
  Phys.\ Rev.\  D {\bf 61}, 036007 (2000)
  [arXiv:hep-ph/9905500].

\bibitem{dawson}

S. Dawson and S. Willenbrock, Phys. Rev. Lett. {\bf 62}, 1232(1989);
W. Marciano, g. Valencia and S. Willenbrock, Phys. Rev. {\bf D40},
1725(1989); L. Durand, J. Johnson and J. Lopez, Phys. Rev. Lett. {\bf 64}, 1215(1990);
Phys. Rev. {\bf D45}, 3112(1992).


\end{references}
\end{document}